\documentclass[fleqn,twoside]{article}
\usepackage{espcrc2}

\newcommand{\prd}{\partial}

\newcommand{\ice}[1]{\relax}

\newcommand{\als}{\alpha_s}
\newcommand{\as}{a_s}

\newcommand{\ovl}{\overline}
\newcommand{\be}{\beta}
\newcommand{\g}{\gamma}

\newcommand{\re}[1]{(\ref{#1})}

\def\beq{\begin{equation}}
\def\eeq{\end{equation}}
\def\bea{\begin{eqnarray}}
\def\eea{\end{eqnarray}}
\def\bq{\begin{quote}}
\def\eq{\end{quote}}

\def\nnb{\nonumber}

\def\nnb{\nonumber}

\def\ba{\begin{array}}
\def\ea{\end{array}}

\newcommand{\BreakI}{ \right. \nonumber \\ &{}& \left. }

\newcommand{\AmS}{{\protect\the\textfont2
  A\kern-.1667em\lower.5ex\hbox{M}\kern-.125emS}}

\title{
\vspace*{-3cm}
{
\centerline{\normalsize\hfill  SFB/CPP-09-52}
\centerline{\normalsize\hfill  TTP09-19   }
}
\vspace{2cm}\boldmath 
$R(s)$ and hadronic $\tau$-Decays in Order $\alpha_s^4$:  technical 
aspects\thanks{
Invited talk given by K.~G.~Ch.  at ``X-th INTERNATIONAL WORKSHOP
                   ON TAU LEPTON PHYSICS'', 22-25 September 2008, Novosibirsk, Russia.
}}

\author{P.~A.~Baikov\address{Institute of Nuclear Physics,
        Moscow State University, \\
        Moscow~~119992, Russia}%
       ,
        K.~G.~Chetyrkin\address[KUNI]{Institut f\"ur Theoretische Teilchenphysik, \\
        Universit\"at Karlsruhe, D-76128 Karlsruhe, Germany
        }%
\thanks{On leave from Institute for Nuclear Research
of the Russian Academy of Sciences, Moscow, 117312, Russia.}
        and
        J.~H.~K\"uhn\addressmark[KUNI]}

\begin{document}

\begin{abstract}
We report on some technical aspects of our calculation of $\alpha_s^4$
corrections to $R(s)$ and the semi-leptonic $\tau$ decay width
\cite{Baikov:2008jh,Baikov:2008cp,Baikov:2004ku}. We discuss the inner structure of the result as
well as the issue of its correctness.  We demonstrate  recently appeared
{\em independent evidence} {\bf positively} testing {\em one} of two
components of our full result.
\vspace{1pc}
\end{abstract}
\maketitle

\section{Introduction}

Three important physical observables, namely, the ratio $ R(s) =
{\sigma(e^+e^-\to {\rm hadrons})\over \sigma(e^+e^-\to \mu^+\mu^-)}\,
$, the hadronic decay rate of the $Z$-boson and 
the semileptonic branching ratio of the
 $\tau$-lepton are
expressed through  the vector and axial-vector current correlators
(see, e. g. reviews \cite{ChKK:Report:1996,Davier:2005xq}).
Perturbative QCD  provides reliable predictions for these
correlators in the continuum, i.e. sufficiently above the respective
quark threshold and the respective resonance region.

The ${\cal O}(\alpha_s^3)$ result for the massless vector
correlator\footnote{ Note that in the massless limit vector and
axial-vector correlators are equal  {\em provided } one
considers only non-singlet contributions and ignores so-called singlet
ones. The latter are absent in the tau-lepton case and  
numerically small for the $Z$-boson decay rate. In the present work we
will discuss non-singlet contributions only.}
has been known since many years \cite{Gorishnii:1991vf,Surguladze:1991tg}.

Recently the calculation of the next, order $\alpha_s^4$
contribution to the vector correlator has been performed
\cite{Baikov:2008jh,Baikov:2008cp,Baikov:2004ku}.  The aim of the present work is to discuss some
technical aspects of our calculations 
as well as to provide some new arguments in favour of their correctness.

Due to lack of space  no phenomenological implications of
\cite{Baikov:2008jh,Baikov:2008cp} are discussed  and the
 interested
reader is referred to Refs.~\cite{Baikov:2008jh,Davier:2008sk,Beneke:2008ad,Maltman:2008nf,Jegerlehner:2008rs,Nesterenko:2008fb}.

\section{Generalities}

Consider  the two-point correlator
of vector   quark currents and the corresponding vacuum polarization
function  ($j_\mu^v= \ovl{Q}\gamma_\mu Q$;
$Q$ is a  quark field with  mass $m$, all other $n_f-1$ quarks  are
assumed to be massless)
\begin{eqnarray}
\nnb
\Pi_{\mu\nu}(q)  &=&
 i \int {\rm d} x e^{iqx}
\langle 0|T[ \;
\;j_{\mu}^{v}(x)j_{\nu}^{v}(0)\;]|0 \rangle
\\
&=&
\displaystyle
(-g_{\mu\nu}q^2  + q_{\mu}q_{\nu} )\Pi(q^2)
{}.
\label{PiV}
\end{eqnarray}
The  physical observable $R(s)$ is related to $\Pi(q^2)$
by
\beq
R(s) = 12 \pi \Im \, \Pi(q^2 + i\epsilon)
{}.
\label{R(S):def}
\eeq
For future reference it is convenient to decompose $R(s)$ into the  massless contribution  
and the one  
quadratic in the quark mass  as follows ($ a_s = \als(\mu^2)/\pi$):
\bea
\nnb
R(s) &=& 3 \left\{r^V_0 +  \frac{m^2}{s} r^{V}_2 \right\}+\dots
\\
 &=& 3\left\{\sum_{i \ge 0} a_s^i\left(
 r^{V,i}_0 + \frac{m^2}{s} r^{V,i}_2 \right)
                                    \right\} +\dots
\nnb
{}.
\eea
The corresponding  representation for the polarization function reads
($Q^2 \equiv -q^2$)
\beq
\Pi = \Pi_0(L, \as) + \frac{m^2}{Q^2} \Pi_2(L,\as)
+ {\cal O }(1/Q^4)
\label{Pi_decomp}
{}.
\eeq
Note that both function on the rhs of  \re{Pi_decomp} depend on only $a_s$ and
$L = \ln\frac{\mu^2}{Q^2}$ and could be conveniently decomposed as follows ($n=0, 2$)
\beq
\Pi_n = 
\,\sum_{i \ge 0} \Pi^n_{i}\,a_s^i,
\ \ 
\Pi^n_i = \sum_{0 \le j \le i+1}\Pi^n_{i,j}\, L^j
{}.
\label{Pin_decomp}
\eeq
The terms in \re{Pin_decomp} without $L$-dependence do not
contribute to $R(s)$.

For the calculation of $r^{V,4}_0$  the divergent
parts of five-loop  and the finite parts of the four-loop 
diagrams are needed 
\cite{ChKT:vv:as2}. The organization of  the calculation
is best based on using 
of the evolution equation for $\Pi$ (see, e.~g. \cite{gvvq})
\beq
\frac{\prd }{\prd L} \Pi_0 =
\g^{VV}(a_s)
-\left(
 \beta( a_s) a_s\frac{\prd }{\prd a_s}
\right) \Pi_0
\label{rgPi0}
{},
\eeq
where
$\g^{VV} = \sum_{i \ge0} \g^{VV}_{i} a_s^i$ is 
the (subtractive) anomalous dimension of the 
correlator \re{PiV} and 
$\beta =  -\sum_{i \ge 0} \beta_i\,  a_s^{(i+1)}$
is the QCD $\beta$-function.

To evaluate  the $L$-dependent  pieces of the  polarization
function  $\Pi^0_{0} \dots \Pi^0_{4}$  in terms of $\g^{VV}_0 \dots \g^{VV}_4$ and $\Pi^0_{0} \dots \Pi^0_3$ 
the evolution eq. \re{rgPi0}  can be solved perturbatively: 
\beq
\ba{l}
\Pi^0_0 = \g^{VV}_0\,L + \Pi^{0}_{0,0}, \ \  \Pi^0_1 =\g^{VV}_1\,L + \Pi^{0}_{1,0},
\ea
\eeq
\beq
\ba{l}
\Pi^0_2 = \be_0\,\g^{VV}_1\,\frac{L^2}{2} + L\,\left(\g^{VV}_2 + \be_0\,\Pi^{0}_{1,0}\right)+ \Pi^{0}_{2,0},
\nnb
\ea
\eeq
\beq
\ba{l}
\Pi^0_3 =\be_0^2\,\g^{VV}_1\,\frac{L^3}{3}
\\
+\frac{L^2}{2}\,\left(\be_1\,\g^{VV}_1+2\,\be_0\,\g^{VV}_2
+ 2\,\be_0^2\,\Pi^{0}_{1,0}\right) 
\\
\phantom{\Pi^0}
+ L\,\left(\g^{VV}_3 + \be_1\,\Pi^{0}_{1,0} + 2\,\be_0\,\Pi^{0}_{2,0}\right)
+\Pi^{0}_{3,0},
\ea
\eeq
\beq
\ba{l}
\Pi^0_4 = \be_0^3\,\g^{VV}_1\,\frac{L^4}{4} + L^3\,\left(\frac{5}{6}\,\be_0\,\be_1\,\g^{VV}_1
+ \be_0^2\,\g^{VV}_2 
\right.
\\
\left.
+ \be_0^3\,\Pi^{0}_{1,0} \right) + L^2\,( \frac{1}{2}\be_2\,\g^{VV}_1 + \be_1\,\g^{VV}_2 + \frac{3}{2}\,\be_0\,\g^{VV}_3 
\\
+ \frac{5}{2}\,\be_0\,\be_1\,\Pi^{0}_{1,0} + 3\,\be_0^2\,\Pi^{0}_{2,0}) 
+L\,(\g^{VV}_4 + \be_2\,\Pi^{0}_{1,0} 
\\
+ 2\,\be_1\,\Pi^{0}_{2,0} + 3\,\be_0\,\Pi^{0}_{3,0}) + \Pi^{0}_{4,0}
{}.
\ea
\eeq

The evolution equation for $\Pi_2$
describing  the $m^2$-corrections 
looks  similar to \re{rgPi0}, namely \cite{Baikov:2004ku,ChetKuhn90}, 
\beq
\frac{\prd }{\prd L} \Pi_2 =
-\left(
2\, \g_m(a_s)
+ 
\beta( a_s) a_s\frac{\prd }{\prd a_s}
\right) \Pi_2
\label{rgPi2}
{},
\eeq
where $\g_m = -\sum_{i \ge 0} \g_m^{(i+1)} a_s^i$ is the quark mass anomalous dimension.

\section{Results}

We refer the reader to \cite{Baikov:2008jh} for a discussion of various theoretical
tools used  to compute $\Pi_0$ to order $a_s^3$  and $\g^{VV}$ to
$a_s^4$.  We  only want to mention here the indispensable role of the
parallel version \cite{Tentyukov:2004hz} of FORM \cite{Vermaseren:2000nd}
and the availability of  large computing resources. 
The  results for $\g^{VV}$ and $\Pi$ are 
given in the next  two subsections.

\subsection{ Five loop anomalous dimension}
\beq
(4\pi)^2\,\g^{VV}_0 = (4\pi)^2\,\g^{VV}_1 = 4,
\label{gVVq01}
\eeq
\beq
(4\pi)^2\,\g^{VV}_2 =   -
\frac{11}{18} n_f
+
 \frac{125}{12}
{},
\label{gVVq2}
\eeq
\begin{eqnarray}
\lefteqn{(4\pi)^2\,\g^{VV}_3 =  
-\frac{77}{972}
 \, n_f^2
{+} \, n_f
\left[
-\frac{707}{216}
-\frac{110}{27}  \,\zeta_{3}
\right]
\nonumber
}
\\
&{+}&
\frac{10487}{432}
+\frac{110}{9}  \,\zeta_{3}
{},
\label{gVVq3}
\end{eqnarray}
%
\begin{eqnarray}
\lefteqn{(4\pi)^2\,\g^{VV}_4 =
\label{gVVq4}
{} \, n_f^3
\left[
\frac{107}{15552}
+\frac{1}{108}  \,\zeta_{3}
\right]
}
\\
&{+}&\!\!\!\!\! \, n_f^2
\left[
-\frac{4729}{31104}
+\frac{3163}{1296}  \,\zeta_{3}
-\frac{55}{72}  \,\zeta_{4}
\right]
\nonumber\\
&{+}&\!\!\!\!\! \, n_f
\left[
-\frac{11785}{648}
-\frac{58625}{864}  \,\zeta_{3}
+\frac{715}{48}  \,\zeta_{4}
+\frac{13325}{432}  \,\zeta_{5}
\right]
\nonumber\\
&{+}&\!\!\!\!\!
\frac{2665349}{41472}
+\frac{182335}{864}  \,\zeta_{3}
-\frac{605}{16}  \,\zeta_{4}
-\frac{31375}{288}  \,\zeta_{5}
{}.
\nonumber
\end{eqnarray}

\subsection{ ${\cal O}(\alpha_s^3)$ polarization operator}
\vspace{5mm}

\beq
(4\pi)^2\,\Pi^{0}_{0,0} =\frac{20}{3}, \ \ (4\pi)^2\,\Pi^{0}_{1,0} =\frac{55}{3} -16  \,\zeta_{3}
{},
\label{Pim0:00-10}
\eeq
\bea
(4\pi)^2\, \Pi^{0}_{2,0} &=& 
\, n_f
\left[
-\frac{3701}{324}
+\frac{76}{9}  \,\zeta_{3}
\right]
\label{Pim0:20}
\\
&{+}&
\frac{41927}{216}
-\frac{1658}{9}  \,\zeta_{3}
+\frac{100}{3}  \,\zeta_{5}
\nonumber
{},
\eea
\begin{eqnarray}
\lefteqn{(4\pi)^2\, \Pi^{0}_{3,0} =   
\nonumber
\, n_f^2
\left[
\frac{196513}{23328}
-\frac{809}{162}  \,\zeta_{3}
-\frac{20}{9}  \,\zeta_{5}
\right]
}
\\
&{+}& \, n_f
\left[
-\frac{1863319}{5184}
+\frac{174421}{648}  \,\zeta_{3}
-\frac{20}{3}  \,\zeta_3^2
\BreakI
\phantom{+ \, n_f }
-\frac{55}{36}  \,\zeta_{4}
+\frac{1090}{27}  \,\zeta_{5}
\right]
\nonumber\\
&{+}&
\left.
\frac{31431599}{10368}
-\frac{624799}{216}  \,\zeta_{3}
+330  \,\zeta_3^2
\BreakI
\phantom{+}
+\frac{55}{12}  \,\zeta_{4}
+\frac{1745}{24}  \,\zeta_{5}
-\frac{665}{9}  \,\zeta_{7}
\right.
{}.
\label{Pim0:30}
\end{eqnarray}

The results for the analytical calculation of   $\Pi^2_{0,0} \dots \Pi^2_{3,0}$   
have  been reported  in \cite{Baikov:2004ku}
while the four-loop   quark anomalous dimension is known from \cite{Chetyrkin:1997dh,Vermaseren:1997fq} 
and three-loop QCD $\beta$-function from \cite{Tarasov:1980au,Larin:1993tp}.
Numerically we find for the polarization operator
\bea
\Pi^0_0 &=& 0.0422172 +0.0253303 \,L,
\label{Pim0:0N}
\\
\Pi^0_1 &=&-0.00569664 + 0.0253303 \,L,
\label{Pim0:1N}
\\
\Pi^0_2 &=& 0.0457538 - 0.0080559 \,n_f, 
\label{Pim0:2N}
\\ 
 &{}& +(0.0502986 - 0.00292047 \,n_f) \,L, 
\nnb
\\   
&{}& + (0.0348292 - 0.00211086 \,n_f) \,L^2 ,
\nnb
\\
\Pi^0_3 &=&
\label{Pim0:3N}
\\
&{}& \hspace{-12mm} 
+0.23570 - 0.033603 \,n_f + 0.0007394 \,n_f^2\nnb
\\
&{}& \hspace{-12mm}+ (0.462093 - 0.10679 \,n_f + 0.0021836 \,n_f^2 ) \,L  
\nnb
\\
&{}&\hspace{-12mm} + (0.219061 - 0.02644 \,n_f + 0.00048674 \,n_f^2) \,L^2  
\nnb
\\
&{}&\hspace{-12mm} +  (0.063854 - 0.0077398 \,n_f + 0.0002345 \, n_f^2 ) \,L^3 
\nnb
{}
\eea
and 
\bea
\Pi^2_0 &=& -0.151982,
\label{Pim2:0N}
\\
\label{Pim2:1N}
\Pi^2_1 &=&-0.405285 - 0.303964 L,
\\
\Pi^2_2 &=& -4.27066 + 0.200532 \,n_f, 
\label{Pim2:2N}
\\ 
 &{}& + (-3.20428 + 0.109765  \,n_f) \,L, 
\nnb
\\   
&{}& + (-0.721913 + 0.0253303  \,n_f) \,L^2 ,
\nnb
\\
\Pi^2_3 &=&
\label{Pim2:3N}
\\
&{}& \hspace{-12mm}
-53.0381 + 5.21239 \,n_f - 0.0740141\, n_f^2
\nnb
\\
&{}& \hspace{-12mm}+ (-43.9568 + 4.05526 \,n_f - 0.0586349 \,n_f^2 ) \,L  
\nnb
\\
&{}&\hspace{-12mm} + (-14.2641 + 1.1082 \,n_f - 0.0182941\, n_f^2) \,L^2  
\nnb
\\
&{}&\hspace{-12mm} + (-1.80478 + 0.143538 \,n_f - 0.00281448\, n_f^2 ) \,L^3 
\nnb
{}.
\eea
Note that at  eqs.~(\ref{gVVq01}-\ref{gVVq3},\ref{Pim0:00-10}-\ref{Pim0:20})
and (\ref{Pim0:0N}-\ref{Pim0:2N},\ref{Pim2:0N}-\ref{Pim2:2N})
as well as $L$-dependent pieces of  of eqs.~(\ref{Pim0:3N},\ref{Pim2:3N}) are, in fact,  known since  
long \cite{Gorishnii:1991vf,Surguladze:1991tg,Gorishnii:1986pz,ChetKuhn90}.  

\section{Final results for $R(s)$}
Our final results for $r^{V,i}_0$ and  $r^{V,i}_2$ are easily obtained from results listed  in 
the  previous two subsections.
Explicit expressions can be found in \cite{Baikov:2008jh,Baikov:2004ku}.

\begin{table}
\begin{center}
\begin{tabular}{|c|c|c|c|c|}
\hline 
 $\ell$  & 1  & 2 & 3 & 4  \\
\hline
&- & $\zeta_3$ & $\zeta_3, \zeta_4, \zeta_5$ &  $\zeta_3, \zeta_4, \zeta_5, \zeta_3^2, \zeta_6, \zeta_7$
\\
\hline
\end{tabular}
\caption{
\label{table1}
Possible irrational structures which are allowed to appear in $\ell$-loop massless propagators.
}
\end{center}
\end{table}

\begin{table}
\begin{center}
\begin{tabular}{|c|c|c|c|c|}
\hline 
 $\ell$  & 1,2  & 3 & 4 & 5  \\
\hline
&- & $\zeta_3$ & $\zeta_3, \zeta_4, \zeta_5$ &  $\zeta_3, \zeta_4, \zeta_5, \zeta_3^2, \zeta_6, \zeta_7$
\\
\hline
\end{tabular}
\caption{
\label{table2}
Possible irrational structures which are allowed to appear in $\ell$-loop anomalous dimensions and $\beta$-functions.
}
\end{center}
\end{table}

It is of interest to discuss the structure of 
trancendentalities appearing in eqs.~(\ref{gVVq01}-\ref{Pim0:30}).
On general grounds one could  expect that  the variety of  $\zeta$-constants entering into
$\ovl{\mbox{MS}}$-renormalized (euclidian) massless propagators\footnote{It is understood that 
${\cal O}(\epsilon^{(5-\ell)}$ terms in an $\ell$-loop  massless propagator could contribute only 
to six-loop anomalous dimension and, thus, are not constrained by Table \ref{table1}.}
  should depend on the loop order according to Table \ref{table1}.
Table \ref{table2} provides  the same information about possible irrational numbers which could show up in 
anomalous dimensions. Table \ref{table2} comes  directly     from Table \ref{table1} by noting  
that  any  $\ell +1$-loop anomalous dimension can be obtained from properly chosen  $\ell$-loop  massless 
propagators  \cite{ChS:R*}.

An examination of eqs.~(\ref{gVVq01}-\ref{Pim0:30}) immediately reveals 
that the real pattern of trancendentalities is significantly more limited 
than what is allowed by Tables \ref{table1} and \ref{table2}. Indeed,  the four-loop anomalous dimension $\g^{VV}_3$ 
contains no $\zeta_4$ and no $\zeta_5$  while the three-loop polarization operator contains  $\zeta_5$ but does not 
comprises $\zeta_4$. The fact of absence of $\zeta_4$ in ${\cal O}(\alpha_s^3)$ contribution to the Adler function 
is well-known  and well-understood  \cite{Gorishnii:1991vf,Broadhurst:1999xk}. 
Why  $\g^{VV}_3$ is free from $\zeta_5$ is still  unclear (at least for us).

Let us  move up one loop. The situation is getting even more intriguing: the five-loop anomalous dimension $\g^{VV}_4$ 
does contain $\zeta_4$ but still does not include  $\zeta_3^2, \zeta_6$ and $ \zeta_7$. The 
four-loop polarization operator contains $\zeta_4$ but is  free from $\zeta_6$. Even more, after we combine
 $\g^{VV}$ and $\Pi^0$ to produce the Adler function, the resulting coefficient in front of $\zeta_4$ happens to be 
zero in a non-trivial way! Indeed, the contribution proportional to $\zeta_4$ from $\Pi^0_{3,0}$ 
reads
\[
\ba{l}
{}3\beta_0\, (\frac{55}{12} - \frac{55}{36}\,n_f) =
{} -\left(-\frac{605}{16} +\frac{715}{48}\,n_f -\frac{55}{72}\,n_f^2\right)
\ea
\nnb
\]
and is  {\em exactly}  opposite in sign to the corresponding piece in \re{gVVq4}! 

Unfortunately, we are not aware about existence of  any ratio behind these  remarkable observations.

\section{How reliable are our results?}

The history of multiloop calculations  teaches us to be cautious. For instance, 
approximately twenty  years ago   a severely wrong result
for the ${\cal O}(\alpha_s^3)$  coefficient  in $R(s)$
was published \cite{Gorishnii:1988bc} and corrected only three years later \cite{Gorishnii:1991vf,Surguladze:1991tg}. 

Now one of these authors (rightfully!) rises an important issue
 of the correctness of the results
\cite{Baikov:2008jh,Baikov:2008cp} and emphasizes the  necessity of
performing their independent test \cite{Kataev:2008nc,Kataev:2008sk}.

We  completely agree with this argumentation.
Unfortunately, at the moment,  we are not aware of 
any independent team which is  going or, at least, able to check our results in full.

However, as described below,  the results of a  recent
calculation  \cite{Hoang:2008qy} 
allow at least for  a (partial) test  of
\cite{Baikov:2008jh,Baikov:2008cp,Baikov:2004ku}.

\subsection{A test of the  polarization operator}

In Ref.~\cite{Hoang:2008qy} a large amount of information about the {\em
massive four-loop polarization} function was collected (its threshold
behavior \cite{Hoang:1998xf,Hoang:1997sj,Hoang:2001mm}, as well as
low-energy moments \cite{Chetyrkin:2006xg,Boughezal:2006px,Maier:2008he} and high-energy
asymptotic \cite{ChetKuhn90,Chetyrkin:1994ex}) in order to restore the
whole function within the Pad\'e approach
\cite{Fleischer:1994ef,Broadhurst:1993mw,Baikov:1995ui} by properly
extending the treatment elaborated more than a decade ago for the 
{\em massive three-loop polarization function}
\cite{Chetyrkin:1996cf}.

Within this  method it is customary to deal with  a ``physically''  normalized  polarization operator
$\hat{\Pi}$ defined such that 
\[
\ba{l}
\hat{\Pi}(M,Q,a_s)
= \Pi(M,Q,a_s)  - \Pi(M,Q=0,a_s)
{},
\ea
\nnb
\]
where $\Pi(M,Q,a_s)$ is defined by eq.\re{Pi_decomp} with  the  use of   the  pole quark mass $M$ mass instead of 
 $\ovl{\mbox{MS}}$ renormalized quark mass $m$ (see \cite{Chetyrkin:1999ys,Chetyrkin:1999qi,Melnikov:2000qh}).
Using the results for  $\Pi(M,Q=0,a_s)$  as listed in  \cite{Chetyrkin:2006xg,Boughezal:2006px} 
we arrive at the following asymptotic behavior of  the (four-loop part of)  $\hat{\Pi}(M,Q,a_s)$ at $Q \to \infty$.
\bea
\hat{\Pi}^0_3 &=& \hat{L}^3\,(-0.0638534543 
\label{Pi03N}
\\
&{}&
\hspace{-10mm} 
 +0.007739812639345\,n_f - 
   0.0002345397769\,n_f^2)
\nnb
\\
&{}& 
\hspace{-10mm} 
+\hat{L}^2\,(0.219061347 - 0.0264409511\,n_f
\nnb
\\
&{}& 
 + 0.000486744424\,n_f^2)
\nnb
\\
&{}& 
\hspace{-10mm}
+\hat{L}\,(-0.4620927910 +  0.1067886396\,n_f
\nnb
\\
&{}& 
0.0021836455422\,n_f^2)
+H_0^{(3)}
\nnb
{},
\eea
\bea
H_0^{(3)} &=&
-11.4121461108 
\label{H03}
\\
&{}&
\hspace{-10mm}
+ 1.4413529302\,n_f 
- 0.032814657849\,n_f^2 
{},
\nnb
\eea
\bea 
\frac{\hat{\Pi}^2_3}{Q^2} &=&
 \frac{\hat{L}^3}{z}\,(-0.4511958959073 
\label{Pi23N}
\\
&{}& 
\hspace{-10mm}
+  0.03588458587\,n_f - 
   0.00070361933\,n_f^2)
\nnb
\\
&{}&
\hspace{-10mm}
+\frac{\hat{L}^2}{z}\,(3.084755098861
\nnb
\\
&{}&
\hspace{-10mm}
- 0.2601632475816\,n_f + 
   0.004573525651\,n_f^2)
\nnb
\\
&{}& 
\hspace{-10mm}
+\frac{\hat{L}}{z}\,(-6.6515904245
\nnb
\\
&{}&
\hspace{-10mm} 
 + 0.78236916962\,n_f - 
   0.01465873606\,n_f^2
) 
\nnb
\\
&{}&
+\frac{H_1^{(3)}}{z}
\nnb
{},
\eea
\bea
H_1^{(3)} &=&
-8.16060818463927
\label{H13}
\\
&{}&
\hspace{-10mm}
+
1.0812664904869\,n_f - 
 0.031095026978\,n_f^2 
{}.
\nnb
\eea

To be  in agreement with  the notations of \cite{Hoang:2008qy} we have used in (\ref{Pi03N},\ref{Pi23N}) 
$ \hat{L} \equiv \ln(Q^2/M^2), \ \ z \equiv -\frac{Q^2}{4\,M^2}$ and  set the renormalization scale $\mu = M$. 
In addition,  we have separated in eqs.~(\ref{Pi03N},\ref{Pi23N}) $\hat{L}$-dependent pieces (known since long)   from
the new $\hat{L}$-independent ones. 

The authors of \cite{Hoang:2008qy} did not have at  their disposal the  $\hat{L}$-independent terms $H_0^{(3)}$ and 
$H_1^{(3)}$ and, thus, did not use them. Instead, they were  able to {\em reconstruct} these terms  for two 
particular values of $n_f = 4,5$ from their final Pad\'e approximants. Table \ref{table3} compares  their (approximate)
results with our exact ones. 
\begin{table}[t]
\begin{center}
\begin{tabular}{|c|c|c|c|}
\hline 
 & & $n_f=4$ & $n_f=5$ \\
\hline
$H_0^{(3)}$ & Pad\'e & $-6.122 \pm 0.054$ & $-4.989 \pm 0.053$
\\
\hline
$H_0^{(3)}$ & exact  & $-6.17176892  $ & $ -5.02574791 $
\\
\hline
$H_1^{(3)}$ & Pad\'e & $-3.885 \pm 0.417 $ & $-3.180 \pm 0.405$
\\
\hline
$H_1^{(3)}$ & exact  & $-4.33306265 $ & $-3.53165141 $
\\
\hline
\end{tabular}
\caption{
\label{table3}
Comparison of the    Pad\'e method predictions for  $H_0^{(3)}$ and $H_1^{(3)}$ 
with exact results.}
\end{center}
\end{table}
We observe the full agreement (within accuracy of the   Pad\'e  approach) between our (exact) results
and approximate ones obtained in \cite{Hoang:2008qy}.

\subsection{Discussion}

The full result for $R(s)$ is composed from two parts: the four-loop polarization operator (including its constant, that is $\ln(Q^2)$   
independent terms) and the five-loop anomalous dimension $\g^{VV}$. At the level of separate Feynman diagrams 
the evaluation of  {\em both} parts is reduced to the direct calculation of four-loop massless propagators \cite{gvvq}. 
On the other hand, the input data used  in  \cite{Hoang:2008qy} came from three different sources:
\\
(i) the threshold behavior of the polarization operator;
\\
(ii) the $\ln(Q^2)$ dependent part of the high energy limit  of the four-loop  polarization operator which technically
was obtained  by  a  calculation of {\em three-loop} massless propagators only;
\\ 
(iii)
the first two physical moments of   the four-loop  polarization operator which technically were obtained 
by a calculation of   massive {\em tadpoles} only. 

Thus, we consider  the full agreement demonstrated  by Table \ref{table3} as a non-trivial  and completely independent 
confirmation of the correctness of the results of \cite{Baikov:2008jh,Baikov:2008cp,Baikov:2004ku}.

\section{Acknowledgments} 

This work was supported by
the Deutsche Forschungsgemeinschaft in the
Sonderforschungsbereich/Transregio
SFB/TR-9 ``Computational Particle Physics''
and by RFBR (grant 08-02-01451).

Our calculation could not be performed without intensive use of the HP
XC4000 super computer of the federal state Baden-W\"urttemberg at the
High Performance Computing Center Stuttgart (HLRS) under the grant
``ParFORM''.

\end{document}